\documentclass[aps,prd,twocolumn,showpacs,amsmath,amssymb,nofootinbib]{revtex4-1}
\usepackage{epsfig}
\usepackage{graphicx}
\usepackage{dcolumn}
\usepackage{bm}
\usepackage{overpic}
\usepackage{subfigure}
\usepackage{float}
\usepackage{color}
\usepackage{amsmath} 
\usepackage{mathcomp}
\usepackage{multirow}
\usepackage{rotating}
\usepackage[colorlinks,linkcolor=blue, citecolor=blue, urlcolor=black]{hyperref}
\newcommand{\PreserveBackslash}[1]{\let\temp=\\#1\let\\=\temp}
\newcolumntype{C}[1]{>{\PreserveBackslash\centering}p{#1}}
\newcolumntype{R}[1]{>{\PreserveBackslash\raggedleft}p{#1}}
\newcolumntype{L}[1]{>{\PreserveBackslash\raggedright}p{#1}}

\begin{document}
\normalsize
\parskip=5pt plus 1pt minus 1pt

\title{ Observation of {\boldmath{$\psi(3686)\to\Xi(1530)^{0}\bar{\Xi}(1530)^{0}$}} and {\boldmath{$\Xi(1530)^{0}\bar{\Xi}^0$}}}

\author{
\begin{small}
\begin{center}
M.~Ablikim$^{1}$, M.~N.~Achasov$^{10,b}$, P.~Adlarson$^{67}$, S.~Ahmed$^{15}$, M.~Albrecht$^{4}$, R.~Aliberti$^{28}$, A.~Amoroso$^{66A,66C}$, M.~R.~An$^{32}$, Q.~An$^{63,49}$, X.~H.~Bai$^{57}$
% ...
\\
\vspace{0.2cm}
(BESIII Collaboration)\\
\vspace{0.2cm} {\it
$^{1}$ Institute of High Energy Physics, Beijing 100049, People's Republic of China\\
% ...
$^{4}$ Bochum Ruhr-University, D-44780 Bochum, Germany\\
% ...
$^{10}$ G.I. Budker Institute of Nuclear Physics SB RAS (BINP), Novosibirsk 630090, Russia\\
% ...
$^{15}$ Helmholtz Institute Mainz, Staudinger Weg 18, D-55099 Mainz, Germany\\
% ...
$^{28}$ Johannes Gutenberg University of Mainz, Johann-Joachim-Becher-Weg 45, D-55099 Mainz, Germany\\
% ...
$^{32}$ Liaoning Normal University, Dalian 116029, People's Republic of China\\
% ...
$^{49}$ State Key Laboratory of Particle Detection and Electronics, Beijing 100049, Hefei 230026, People's Republic of China\\
% ...
$^{57}$ University of Jinan, Jinan 250022, People's Republic of China\\
% ...
$^{63}$ University of Science and Technology of China, Hefei 230026, People's Republic of China\\
% ...
$^{66}$ University of Turin and INFN, (A)University of Turin, I-10125, Turin, Italy; (B)University of Eastern Piedmont, I-15121, Alessandria, Italy; (C)INFN, I-10125, Turin, Italy\\
$^{67}$ Uppsala University, Box 516, SE-75120 Uppsala, Sweden\\
% ...
$^{b}$ Also at the Novosibirsk State University, Novosibirsk, 630090, Russia\\
% ...
}\end{center}
\vspace{0.4cm}
\end{small}
}


\vspace{4cm}
\date{\today}

\begin{abstract}
 Using $(448.1\pm2.9)\times 10^{6}$ $\psi(3686)$ events collected with
 the BESIII detector and a single-baryon tagging technique,
  we present the first observation of the decays $\psi(3686)\to\Xi(1530)^{0}\bar{\Xi}(1530)^{0}$ and $\Xi(1530)^{0}\bar{\Xi}^0$.
 The branching fractions are measured to be
 ${\cal{B}}(\psi(3686)\to\Xi(1530)^{0}\bar{\Xi}(1530)^{0}) = (6.77\pm0.14\pm0.39)\times10^{-5}$ and
${\cal{B}}(\psi(3686)\to\Xi(1530)^{0}\bar{\Xi}^{0}) = (0.53\pm0.04\pm0.03)\times10^{-5}$.
Here, the first and second uncertainties are statistical and systematic, respectively.
In addition, the parameter associated with the angular distribution for the decay $\psi(3686)\to\Xi(1530)^{0}\bar{\Xi}(1530)^0$ is determined to be $\alpha = 0.32\pm0.19\pm0.07$, in agreement with theoretical predictions within one standard deviation.

\end{abstract}

\maketitle

\section{Introduction}
\label{sec:introduction}
\vspace{-0.4cm}
The production of the $J/\psi$ and $\psi(3686)$ resonances (here both denoted by $\psi$) in $e^{+}e^{-}$ annihilation and their subsequent two-body hadronic decays can be used to test perturbative Quantum Chromodynamics (QCD) and QCD-based calculations~\cite{ref1}, which concern topics such as gluon spin, quark-distribution amplitudes in baryon-antibaryon ($B\bar{B}$) pairs, and total hadron-helicity conservation. In particular, the decay of charmonium to $B\bar{B}$ pairs is a suitable avenue for testing these calculations due to its simple topology. However, only a few $\psi\to B\bar{B}$ decay modes have been precisely measured until now. Many decay modes, such as $\psi(3686)\to\Xi(1530)^{0}\bar{\Xi}(1530)^{0}$ and $\Xi(1530)^{0}\bar{\Xi}^0$ decays, are still either unknown or less precisely measured~\cite{cleo1530}.

A precise measurement of the angular distributions of $\psi\to B\bar{B}$ is desirable to test helicity conservation rules and theoretical models~\cite{ref3,ref4}, especially the color-octet contributions in these decays. The angular distribution for the process $\psi\to B\bar{B}$ can be expressed as
\begin{equation}
\frac{dN}{d(\cos\theta_{\rm B})}\propto 1+\alpha\cos^2\theta_{\rm B},
\label{eq1}
\end{equation}
where $\theta_{\rm B}$ is the angle of the momentum vector of one of the baryons in the center-of-mass (CM) system with respect to the direction of the positron beam.
Theoretically, the values of the angular-distribution parameter $\alpha$ have been predicted by first-order QCD calculations~\cite{ref3,ref4}.
In the prediction of Claudson $et~al.$~\cite{ref3}, the mass of the final baryon is taken into consideration unitarily, while the constituent quarks inside the baryon are taken as massless when computing the decay amplitude. Yet Carimalo $et~al.$~\cite{ref4} deemed that quark mass effects are more sensitive than electromagnetic contributions to the $\alpha$ value
, and the mass effects at the quark level are taken into account.
Experimentally, the values of $\alpha$ have been measured for several baryon-pair final states~\cite{ref5,ref6,ref7,ref8,ref10}.
Positive values of $\alpha$ were found for the processes $\psi(3686)\to\Xi(1530)^{-}\bar{\Xi}(1530)^{+}$~\cite{memowxf} and $\psi(3686)\to\Xi^0\bar\Xi^0$, while for other known processes, such as $J/\psi\to\Sigma^0\bar\Sigma^0$ and $\Sigma(1385)\bar\Sigma(1385)$, the values were found to be negative. Chen and Ping~\cite{Ping} noted that the angular distribution parameter for $\psi\to B\bar{B}$ could be negative when rescattering effects of $B\bar{B}$ in heavy quarkonium decays are taken into account.
The BESIII experiment has collected a large data sample at the $\psi(3686)$ peak, which can be used to investigate several baryon decay channels {\it e.g.}, $\psi(3686)\to\Xi(1530)^{0}\bar{\Xi}(1530)^{0}$, where the $\alpha$ value is predicted to be
 0.32 by C. Carimalo~\cite{ref4}.

An additional motivation for studying the baryonic decays of $\psi$ mesons is that perturbative QCD predicts that the ratio of the branching fractions of the $\psi(3686)$ and the $J/\psi$ decaying into the same final state should follow the so-called `12\% rule'~\cite{add1}:
\begin{equation}
   {\cal{Q}}=\frac{{\cal{B}}(\psi(3686)\to \rm hadrons)}{{\cal{B}}(J/\psi\to \rm hadrons)}\approx 12\%.
\end{equation}
Although the ratio has been measured with a wide variety of final states~\cite{pdg}, the review in Ref.~\cite{add2}, which includes a comparison of theory and experiment, concludes
that the current theoretical explanations are unsatisfactory.
More experimental results, such as the decay of $\psi(3686)\to \Xi(1530)^{0}\bar{\Xi}^0$, are required to
test the `12\% rule'~\cite{add1} and theoretical models~\cite{add2}.

Finally, according to SU(3)-flavor symmetry, decays of charmonium into $B\bar{B}$ pairs of the same multiplet ($B_{1}\bar{B}_{1}$, $B_{8}\bar{B}_{8}$ and $B_{10}\bar{B}_{10}$, where $B_1$, $B_8$ and $B_{10}$ represent SU(3) singlet, octet, and decuplet baryons, respectively) are allowed,
but those into octet-decuplet baryon pairs $B_{8}\bar{B}_{10}$ are forbidden~\cite{ref20}.
Violations of these predictions have long been known.
For instance,  $J/\psi\to B_{8}\bar{B}_{10}$ including $J/\psi\to\Xi^0\bar{\Xi}(1530)^0$, $\Xi^-\bar{\Xi}(1530)^+$, $\bar\Sigma^{+}\Sigma(1385)^{-}$, $\bar\Sigma^{-}\Sigma(1385)^{+}$ {\it etc.}, were reported by the DM2 Collaboration~\cite{ref200}. Recently, the BESIII Collaboration has reported an improved measurement of $J/\psi\to\Xi^{-}\bar{\Xi}(1530)^{+}$~\cite{refjiqp} and the first observation of $\psi(3686)\to\Xi^{-}\bar{\Xi}(1530)^{+}$~\cite{memowxf}.  Additional measurements on this topic are desirable.

In this paper, branching fraction measurements for  charmonium decays to the neutral $B_{10}\bar{B}_{10}$ and $B_{10}\bar{B}_{8}$ modes, $\psi(3686)\to\Xi(1530)^{0}\bar{\Xi}(1530)^{0}$ and $\psi(3686)\to\Xi(1530)^{0}\bar{\Xi}^0 $, as well as the angular distribution measurement for $\psi(3686)\to\Xi(1530)^{0}\bar{\Xi}(1530)^{0}$, are presented using $(448.1\pm2.9)\times 10^{6}$ $\psi(3686)$ events collected with the BESIII detector at BEPCII in 2009 and 2012~\cite{Npsip}.
To increase the detection efficiency, as in Ref.~\cite{wxfpaper}, this analysis employs a single-baryon tagging technique in which we reconstruct a $\Xi(1530)^{0}$ candidate through the decay chain $\Xi(1530)^0\to\Xi^-\pi^+$,  with $\Xi^-\to\Lambda\pi^-$ and $\Lambda\to p\pi^-$,  and search for a  $\bar\Xi(1530)$ ($\bar\Xi^{0}$) candidate on the recoil side in $\psi(3686)\to\Xi(1530)^{0}\bar\Xi(1530)^{0}$ ($\psi(3686)\to\Xi(1530)^{0}\bar\Xi^{0}$) decay (unless otherwise noted, charge conjugation is implied throughout the paper).

\section{BESIII detector and monte carlo simulation}
\label{sec:BESIII}
\vspace{-0.4cm}
The BESIII detector~\cite{Ablikim:2009aa} records symmetric $e^+e^-$ collisions
provided by the BEPCII storage ring~\cite{Yu:IPAC2016-TUYA01}, which operates with a peak luminosity of $1\times10^{33}$~cm$^{-2}$s$^{-1}$
in the center-of-mass energy range from 2.0 to 4.9~GeV.
The cylindrical core of the BESIII detector covers 93\% of the full solid angle and consists of a helium-based
 multilayer drift chamber~(MDC), a plastic scintillator time-of-flight
system~(TOF), and a CsI(Tl) electromagnetic calorimeter~(EMC),
which are all enclosed in a superconducting solenoidal magnet
providing a 1.0~T
magnetic field. The solenoid is supported by an
octagonal flux-return yoke with resistive plate counter muon
identification modules interleaved with steel.
The charged-particle momentum resolution at $1~{\rm GeV}/c$ is
$0.5\%$, and the specific energy loss (d$E$/d$x$) resolution is $6\%$ for electrons
from Bhabha scattering. The EMC measures photon energies with a
resolution of $2.5\%$ ($5\%$) at $1$~GeV in the barrel (end-cap)
region. The time resolution in the TOF barrel region is 68~ps, while
that in the end-cap region is 110~ps.

Simulated data samples produced with a {\sc
geant4}-based~\cite{geant4} Monte Carlo (MC) package, which
includes the geometric description of the BESIII detector and the
detector response, are used to determine detection efficiencies
and to estimate background contributions. The simulation models the beam-energy spread and the initial state radiation (ISR) in the $e^+e^-$
annihilations with the generator {\sc
kkmc}~\cite{ref:kkmc}. The inclusive MC sample includes the production of the
$\psi(3686)$ resonance, the ISR production of the $J/\psi$, and
the continuum processes incorporated in {\sc
kkmc}~\cite{ref:kkmc}.
The known decay modes are modelled with {\sc
evtgen}~\cite{ref:evtgen} using branching fractions taken from the
Particle Data Group (PDG)~\cite{pdg}, and the remaining unknown charmonium decays
are modelled with {\sc lundcharm}~\cite{ref:lundcharm}. Final-state radiation~(FSR)
from charged final state particles is incorporated using the {\sc
photos} package~\cite{photos}.
To determine the detection efficiencies for the signal modes $\psi(3686)\to\Xi(1530)^{0}\bar{\Xi}(1530)^{0}$ and $\psi(3686)\to\Xi(1530)^{0}\bar{\Xi}^0 $,
1 million simulated signal MC samples are generated for each reconstructed mode, taking
into account the angular distributions measured in this analysis.
The processes $\Xi(1530)^0\to\Xi^-\pi^+$ with $\Xi^-\to\Lambda\pi^-$ and $\Lambda\to p\pi^-$ are generated uniformly in phase space.
The data set collected at the CM energy of 3.65 (3.572) GeV with an integrated luminosity of 43.88 (23.14) $\rm pb^{-1}$
is used to estimate the contamination
from the continuum processes $e^+e^-\to\Xi(1530)^{0}\bar{\Xi}(1530)^{0}$ and $e^+e^-\to\Xi(1530)^{0}\bar{\Xi}^0 $~\cite{Npsip}.

\section{Data Analysis}
\label{sec:selection}
Charged tracks detected in the MDC are required to be within a polar angle ($\theta$) range of $|\cos\theta|<0.93$, where $\theta$ is defined with respect to the axis of the MDC.
Particle identification~(PID) for charged tracks combines measurements of the specific energy loss
d$E$/d$x$ in the MDC and the flight time measured by the TOF to form likelihoods $\mathcal{L}_h~(h=p,K,\pi)$ for each hadron $h$ hypothesis.
Tracks are identified as protons (pions) when the proton (pion) hypothesis has the greatest likelihood, i.e. $\mathcal{L}_{p(\pi)}>\mathcal{L}_{K}$ and $\mathcal{L}_{p(\pi)}>\mathcal{L}_{\pi(p)}$.

The $\Lambda$ candidates are reconstructed from $p\pi^{-}$ pairs with an invariant mass lying within $\pm 5~\text{MeV}/c^2$ of the nominal $\Lambda$ mass~\cite{pdg}. This mass window is chosen by optimizing the figure-of-merit (FOM) $\frac{S}{\sqrt{S+B}}$, where $S$ is the number of signal events and $B$ is the number of background events. In this case, $S$ and $B$ are taken
from signal MC and the inclusive MC samples, respectively.
A secondary-vertex fit~\cite{secondvt} is performed with all the $p\pi^{-}$ combinations; those with
$\chi^{2}< 500$ are kept for further analysis. To further suppress
background events, the decay length of the $\Lambda$ candidate is required to be
positive. In the case of multiple candidates, the one with an
unconstrained mass closest to the nominal mass is retained.
The $\Xi^-$ candidates are reconstructed from $\Lambda\pi^{-}$ pairs with an invariant mass within $\pm 8~\text{MeV}/c^2$ of the nominal $\Xi^-$ mass~\cite{pdg}.
A secondary-vertex fit~\cite{secondvt} is performed on all $\Lambda\pi^{-}$ combinations. Only the candidate with an invariant mass closest to the nominal mass is retained if there is
more than one candidate in an event. The decay length of the $\Xi^-$ is required to be positive to further suppress background. The $\Xi(1530)^0$ candidates are
reconstructed by combining $\Xi^-$ candidates with an additional $\pi^+$, minimizing the variable $|M_{\Xi\pi}-M_{{\Xi(1530)}^0}|$, where $M_{\Xi\pi}$ is the invariant mass of $\Xi^-\pi^+$ and $M_{{\Xi(1530)}^0}$ is the nominal mass of the $\Xi(1530)^0$~\cite{pdg}.  On average, there are 1.07 (1.02) candidates per event for the $\Xi(1530)^0 \bar \Xi(1530)^0$ ($\Xi(1530)^0 \bar \Xi^0$) final state before the best candidate selection.

A further requirement of $|M^{\rm recoil}_{\pi^+\pi^-}-M_{J/\psi}|>4~\text{MeV}/c^2$, obtained by optimizing the FOM, is applied to suppress the $\psi(3686)\to\pi^+\pi^- J/\psi$ background, where the $M^{\rm recoil}_{\pi^+\pi^-}$ is the
recoil mass for the $\pi^+\pi^-$ system, and $M_{J/\psi}$ is the nominal mass of the $J/\psi$ meson~\cite{pdg}.

\begin{figure}[htbp]
\centering
  \mbox{
    \begin{overpic}[width=0.25\textwidth,clip=true]{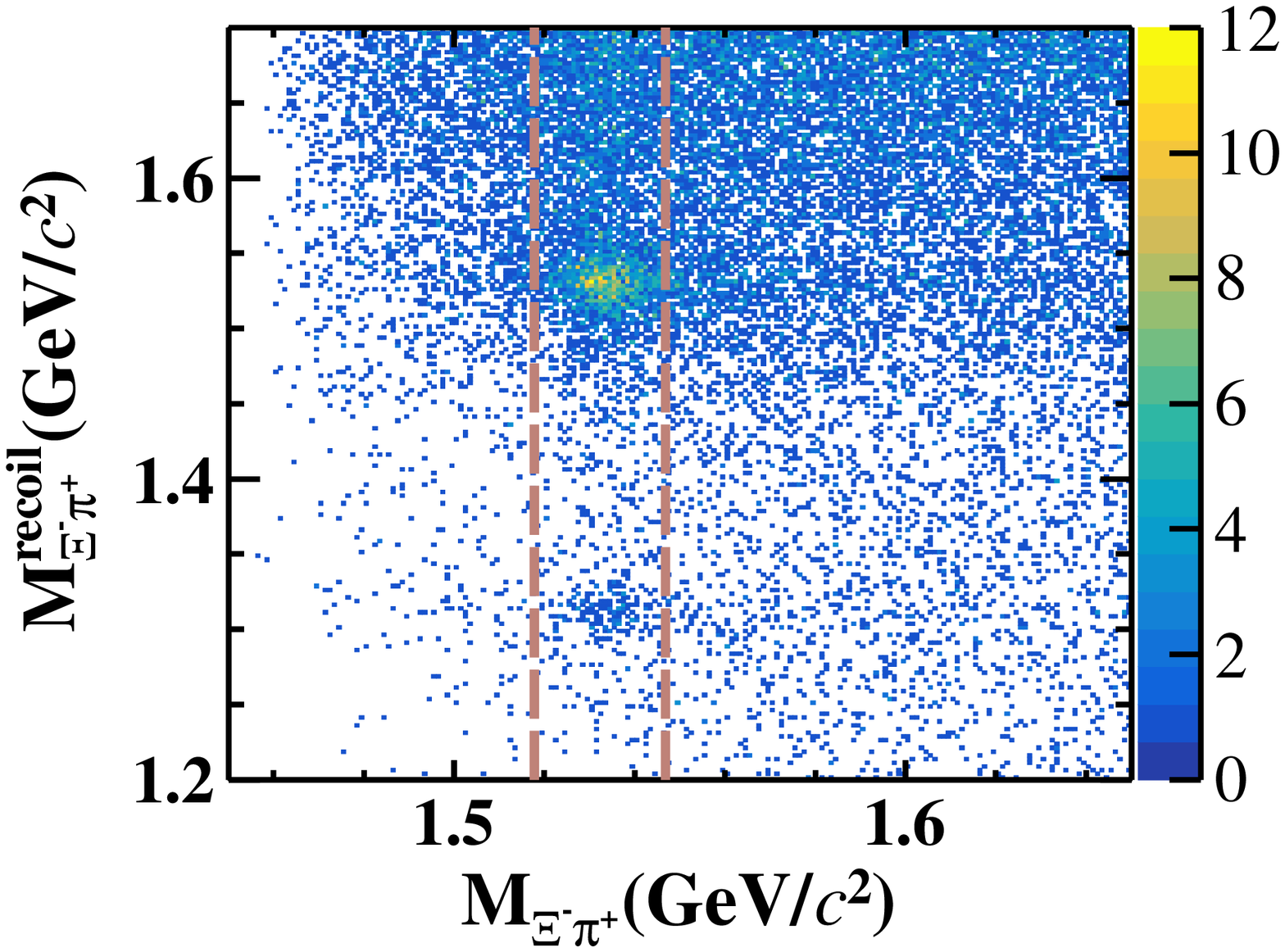}
    \end{overpic}
    \begin{overpic}[width=0.25\textwidth,clip=true]{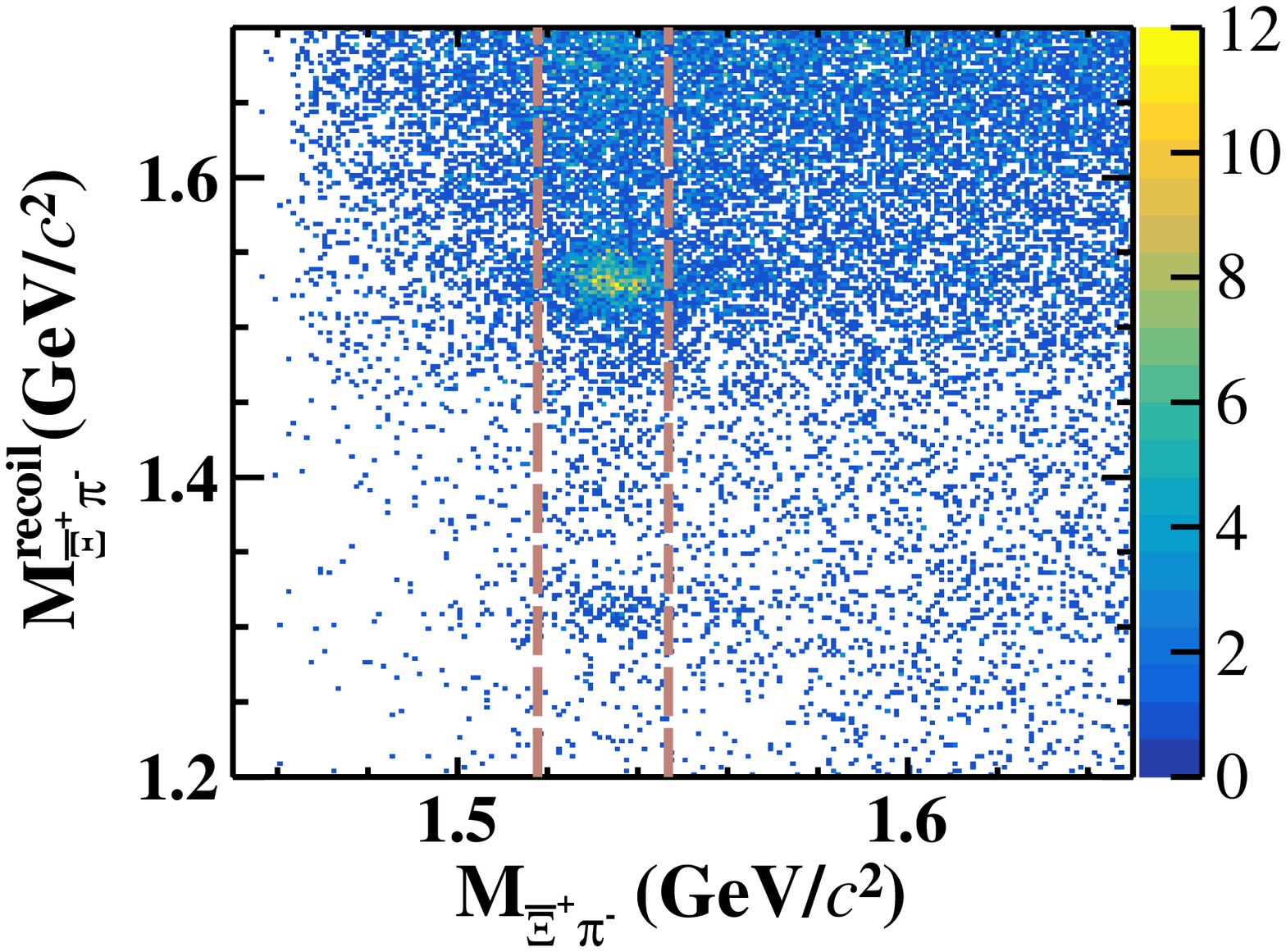}
    \end{overpic}
   }
\caption{Distributions of $M^{\rm recoil}_{\Xi\pi}$ versus $M_{\Xi\pi}$
for $\Xi(1530)^0$ tag (left) and $\bar\Xi(1530)^0$ tag (right). Two dashed lines denote the
$\Xi(1530)^0$ signal region.}
\label{scatter}
\end{figure}

The presence of the anti-baryon candidates $\bar{\Xi}^{0}$ and $\bar{\Xi}(1530)^{0}$ is inferred
from the mass recoiling against the tagged $\Xi^{-}\pi^{+}$ system,
$M^{\rm recoil}_{\Xi^-\pi^+}=\sqrt{(E_{\rm CM}-E_{\Xi^-\pi^+})^2-(\bm{P}_{\Xi^-\pi^+})^2 }$, where $E_{\rm CM}$ is CM energy, $E_{\Xi^-\pi^+}$ and $\bm{P}_{\Xi^-\pi^+}$ are the energy and momentum of the selected $\Xi^-\pi^+$ system in the $e^+e^-$ rest frame. Figure~\ref{scatter} shows the distributions of $M^{\rm recoil}_{\Xi\pi}$ versus $M_{\Xi\pi}$. Clear $\psi(3686)\to\Xi(1530)^{0}\bar{\Xi}(1530)^{0}$ and $\boldmath{\Xi(1530)^{0}\bar{\Xi}^0}$ signals are
observed.

Analysis of the $\psi(3686)$ inclusive MC sample with a generic event-type investigation tool, TopoAna~\cite{topo}, indicates
that the main background events come from $\psi(3686)\to\pi^+\pi^-(\pi^0\pi^0)J/\psi$ with $J/\psi\to\Xi^-\bar\Xi^+$.
The background shape from the $\pi^0\pi^0 J/\psi$ channel is flat, while the background shape from the $\pi^+\pi^- J/\psi$ channel
is not smooth.
We also investigate the contribution from continuum processes with the off-peak data at $\sqrt{s}= 3.65$ (3.572)~GeV. It is found that few events survive the selection criteria, and no $\Xi(1530)^0$ ($\Xi^0$) peaking contribution is seen due to the rather low sample size.
Taking into account the normalization of the integrated luminosity and CM energy dependence of the cross section, the contribution from continuum processes is expected to be
small and is neglected in the remainder of the analysis.
In addition,
there is the possibility of miscombinations of negative pions with similar momenta in both the baryon and anti-baryon decay chains in the signal events. Including these in the $\Xi(1530)^0$ or $\Xi^0$ reconstruction leads to a wrong-combination background (WCB).

\section{Branching Fraction and Angular Distribution Measurements}

\subsection{Branching fractions}
\label{sec:signal}
\begin{figure*}[hbtp]
\centering
  \mbox{
        \begin{overpic}[width=0.90\textwidth,clip=true]{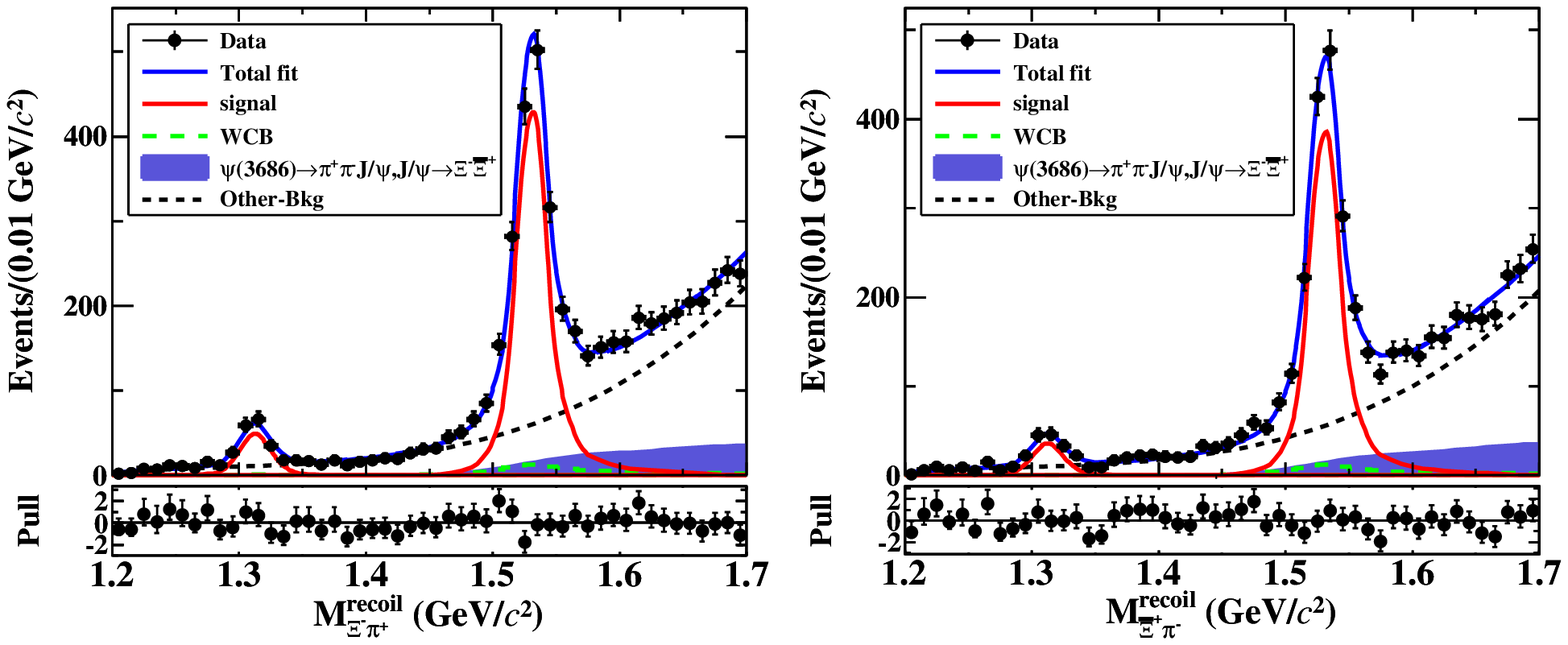}
        \end{overpic}
       }
\caption{Simultaneous fits to the $M^{\rm recoil}_{\Xi^-\pi^+}$ (left) and $M^{\rm recoil}_{\bar\Xi^+\pi^-}$ (right) spectra for $\psi(3686)\to\Xi(1530)^0\bar\Xi(1530)^0$ and $\Xi(1530)^0\bar\Xi^0$ decays, respectively.
Dots with error bars are data.
The solid blue lines represent the fit results, the solid red lines denote the signals,
the dashed green lines denote the wrong-combination background events (WCB), the shaded blue histograms denote the background $\psi(3686)\to\pi^+\pi^- J/\psi$ with $J/\psi\to\Xi^-\bar\Xi^+$, and the short-dashed black lines denote contributions from other background sources (Other-Bkg).
  }
\label{scatter2}
\end{figure*}

The signal yields for the two decays $\psi(3686)\to\Xi(1530)^{0}\bar{\Xi}(1530)^{0}$ and $\Xi(1530)^{0}\bar{\Xi}^0 $ are determined by performing a simultaneous extended unbinned maximum likelihood fit to the $M^{\rm recoil}_{\Xi^-\pi^+}$ and $M^{\rm recoil}_{\bar{\Xi}^+\pi^-}$ spectra, respectively. In the fit, the signal shapes are represented by the
MC-simulated shape sampled from a multidimensional histogram using the signal MC simulation, convolved with a Gaussian function to take into account the detector-resolution difference between data and MC simulation, where the parameters of the Gaussian function are left free but constrained to be the same for the two charge conjugation modes.
The background events from $\psi(3686)\to\pi^+\pi^-J/\psi$ with $J/\psi\to\Xi^-\bar\Xi^+$
are described by their respective MC-simulated shapes and the corresponding number of background events, computed according to their respective branching fractions, are fixed.
The WCB is also described by the MC-simulated shape, and its contribution is fixed according to the MC simulation.
The shape of the other sources of the background (Other-Bkg) looks smooth and is described by a third-order Chebychev polynomial function.
The fit results are shown in Fig.~\ref{scatter2}. The statistical significance for both $\psi(3686)\to\Xi(1530)^0\bar\Xi(1530)^0$ and  $\psi(3686)\to\Xi(1530)^0\bar\Xi^0$ is found to be greater than $10\sigma$.

The branching fraction of $\psi(3686)\to X$ is calculated as
\begin{equation}
          {\cal{B}}(\psi(3686)\to X) = \frac{N^{\rm sig}}
          { N_{\psi(3686)}\cdot {\cal{B}}\cdot\epsilon},
\end{equation}
where $X$ stands for $\Xi(1530)^0\bar{\Xi}(1530)^0$ or $\Xi(1530)^0\bar{\Xi}^0$; $N^{\rm sig}$ is the number of observed events extracted from the fit; $N_{\psi(3686)}$ is the total number of $\psi(3686)$ events~\cite{Npsip}; ${\cal{B}}$ denotes the product of branching fractions of $\Xi(1530)^0\to\Xi^-\pi^+$, $\Xi^-\to\Lambda\pi^-$ and $\Lambda\to p\pi^-$~\cite{pdg}; $\epsilon$ denotes the detection efficiency
obtained from the signal MC sample generated with the measured $\alpha$ value for $\psi(3686)\to\Xi(1530)^{0}\bar{\Xi}(1530)^{0}$ as discussed below.

The signal yields ($N^{\rm sig}$), the detection efficiencies ($\epsilon$) and the branching fractions (${\cal{B}}(\psi(3686)\to X)$) for the analyzed decay modes are summarized in Table~\ref{tab:st0}. The systematic uncertainties on the measurements for the branching fractions will be discussed in Section~\ref{sec:errBR}.

\begin{table*}[htbp]
\caption{Summary of the number of extracted signal yield ($N^{\rm sig}$), efficiency ($\epsilon(\%)$),
the angular distribution parameter ($\alpha$) and branching fractions (${\cal{B}}$) measured in this analysis which are described in detail in Section~\ref{sec:signal} and ~\ref{sec:signal2}, respectively, the combined branching fractions for neutral (charged~\cite{memowxf}) (${\cal{B}}^{\rm com}_{\rm neutral}$ (${\cal{B}}^{\rm com}_{\rm charged}$)), the combined angular distribution parameter for neutral (charged~\cite{memowxf}) mode ($\alpha^{\rm com}_{\rm neutral}$ ($\alpha^{\rm com}_{\rm charged}$)), as well as the ratio of branching fractions between the neutral and charged modes ${\cal{R}}={{\cal{B}}^{\rm com}_{\rm neutral}}/{{\cal{B}}^{\rm com}_{\rm charged}}$. Here, the uncertainties for $N^{\rm sig}$ and $\epsilon$ are statistical only, while for other quantities the first uncertainties are statistical and the second ones are systematics.  `...' denote unmeasured quantities.}
 \begin{center}
  \doublerulesep 2pt\small{
    \newcommand\ST{\rule[-0.75em]{0pt}{1.75em}}

    \begin{tabular}{lcccc}\hline\hline
            & \multicolumn{2}{c}{ $\psi(3686)\to\Xi(1530)^0\bar{\Xi}(1530)^0$} &  \multicolumn{2}{c}{$\psi(3686)\to\Xi(1530)^0\bar{\Xi}^0$} \ST \\
\hline
Tag mode            &$\Xi(1530)^0$ &  $\bar{\Xi}(1530)^0$ & $\Xi(1530)^0$ &  $\bar{\Xi}(1530)^0$ \ST \\\hline
$N^{\rm sig}$           &    $1553\pm50$          &      $1398\pm47$                &     $160\pm16$          &      $118\pm14$               \\
$\epsilon(\%)$      &     $11.76\pm0.03$         &    $11.26\pm0.03$                  &     $13.96\pm0.04$          &     $13.29\pm0.04$                \\
$\alpha$            &    $0.47\pm0.27\pm0.11$          &    $0.20\pm0.25\pm0.10$                  &     ...          &        ...             \\
${\cal{B}}(10^{-5})$&  $6.92\pm0.22\pm0.44$            &    $6.51\pm0.22\pm0.53$                  &    $0.60\pm0.06\pm0.04$           &      $0.47\pm0.06\pm0.04$  \\\hline
$\alpha^{\rm com}_{\rm neutral}$  & \multicolumn{2}{c}{ $0.32\pm0.19\pm0.07$} & \multicolumn{2}{c}{...} \ST\\
$\alpha^{\rm com}_{\rm charged}$~\cite{memowxf}  & \multicolumn{2}{c}{ $0.40\pm0.24\pm0.06$} & \multicolumn{2}{c}{...} \ST\\
${\cal{B}}^{\rm com}_{\rm neutral}(10^{-5})$  & \multicolumn{2}{c}{ $6.77\pm0.14\pm0.39$} & \multicolumn{2}{c}{$0.53\pm0.04\pm0.03$} \ST \\
${\cal{B}}^{\rm com}_{\rm charged}(10^{-5})$~\cite{memowxf}  & \multicolumn{2}{c}{ $11.45\pm0.49\pm0.92$} & \multicolumn{2}{c}{$0.70\pm0.11\pm0.04$} \ST\\
${\cal{R}}$  & \multicolumn{2}{c}{ $0.59\pm0.03\pm0.06$} & \multicolumn{2}{c}{$0.76\pm0.13\pm0.06$} \ST\\
\hline
\hline
\end{tabular}}
\label{tab:st0}
 \end{center}
 \end{table*}

\subsection{Angular distribution}\label{sec:signal2}
The angular-distribution parameter $\alpha$ for the decay $\psi(3686)\to\Xi(1530)^{0}\bar{\Xi}(1530)^{0}$ is determined by performing a least-squares fit based on Eq.~(\ref{eq1}) to the efficiency-corrected distribution of $N^{\rm sig}$ versus $\cos \theta_{\rm B}$. The signal yields are extracted in eight equal bins of $\cos\theta_{\rm B}$ in the range from $-0.8$ to 0.8 using the method described in Section~\ref{sec:signal}. The corresponding signal yield in each bin of $\cos \theta_{\rm B}$ is corrected with an efficiency obtained from a signal MC sample generated uniformly in phase space.

\begin{figure}[hbtp]
\centering
  \mbox{
         \begin{overpic}[width=0.25\textwidth,clip=true]{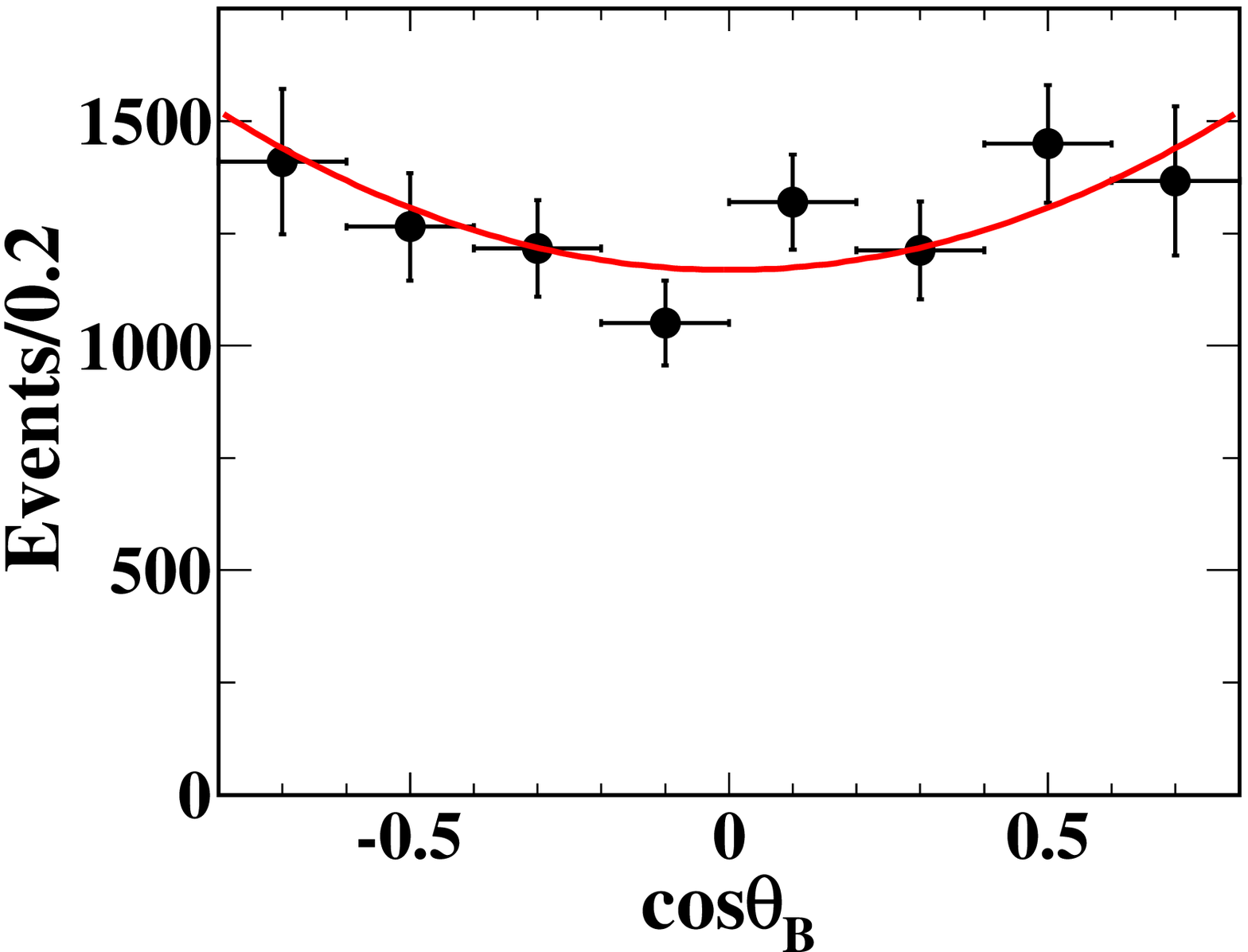}
         \end{overpic}
         \begin{overpic}[width=0.25\textwidth,clip=true]{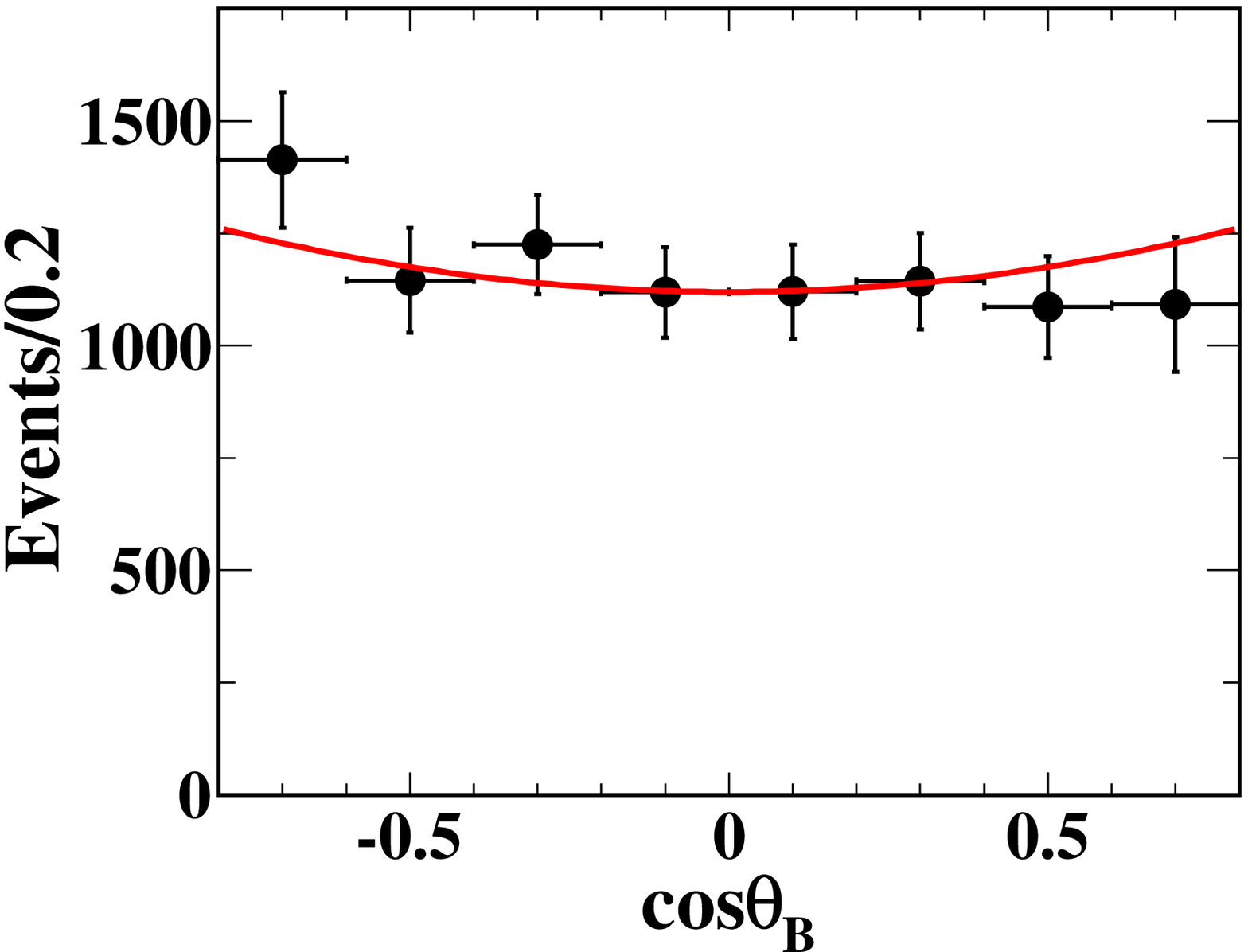}
         \end{overpic}
       }
\caption{Distributions of $\cos\theta_{\rm B}$ for the $\Xi(1530)^0$ tag (left)
and the $\bar\Xi(1530)^0$ tag (right). The dots with error bars
show the efficiency-corrected $N^{\rm sig}$ for data, without including the systematic uncertainty, and the curves show the fit
results.}
\label{angular}
\end{figure}

The distribution of the efficiency-corrected signal yield from data together with the results of the fit are shown in Fig.~\ref{angular}, obtained by separate fits to the corresponding distributions. The goodness of the fit is calculated to be $\chi^2/\rm NDF = 3.58/6$ for the $\Xi(1530)^0$ tag, and $4.82/6$ for the $\bar\Xi(1530)^0$ tag, where NDF is the number of degrees of freedom.
The values of $\alpha$ obtained from the fit are summarized in Table~\ref{tab:st0}. The systematic uncertainty on the measurement of $\alpha$ will be discussed in
Sect.~\ref{sec:errang}.
Due to the limited sample size, $\alpha$ for the decay $\psi(3686)\to\Xi(1530)^{0}\bar{\Xi}^{0}$ is not measured in this study.

\section{Systematic Uncertainties}
\label{sec:syserr}
\subsection{Branching fractions}
\label{sec:errBR}
The systematic uncertainties in the branching fraction measurements are associated with knowledge of the tracking efficiency, PID efficiency, trigger
efficiency, $\Xi^{-}$ reconstruction efficiency, the choice of mass windows and decay lengths for the $\Lambda$ and $\Xi^{-}$ candidates, fit range, signal and background shapes, the knowledge of the branching fraction of the intermediate decay, and the uncertainty on the total number of $\psi(3686)$ events. The uncertainties due to the tracking and PID efficiency (except for those associated with the $\pi^+$  from the $\Xi(1530)^0$ decay) and the $\Lambda$ reconstruction efficiency have been incorporated in the $\Xi^{-}$ reconstruction efficiency, and the uncertainty due to the trigger efficiency is negligible~\cite{trigger}.

Below is a summary of all the sources of the systematic uncertainties identified in this analysis.
\begin{itemize}
\item  {\bf Tracking and PID}: The uncertainties due to the $\pi^+$ tracking (coming from the $\Xi(1530)^{0}$ mother particle) efficiency and the PID efficiency are investigated with a control sample of $\psi(3686)\to \pi^{+}\pi^{-}p\bar{p}$ decays. The observed difference in efficiency between the data and MC simulation, 1.0\%, for both quantities, is assigned as the uncertainty for the tracking and PID efficiency separately~\cite{memowxf}.

\item {\bf $\Xi^-$ reconstruction}: The uncertainty due to the $\Xi^{-}$ reconstruction efficiency is estimated from a study of a control sample of $\psi(3686)\to\Xi^{-}\bar{\Xi}^{+}$ decays~\cite{memowxf}. The efficiency differences between the data and MC simulation are found to be 3.0\% for the $\Xi^-$ reconstruction and 6.0\% for the $\bar{\Xi}^+$ reconstruction, which are taken as the systematic uncertainties.

\item {\bf Decay length of \boldmath $\Lambda$ ($\Xi^-$)}: The uncertainty associated with the efficiency for the decay-length requirement on $\Lambda$ ($\Xi^-$) candidates is estimated as the difference of selection efficiency on the data and MC simulation with a control sample $J/\psi\to\Xi^{-}\bar{\Xi}^{+}$ decays.

\item {\bf Mass window of \boldmath $\Lambda$ ($\Xi^-$)}: The uncertainty associated with the mass-window requirement on the $\Lambda$ ($\Xi^-$) candidates is estimated as the difference of the selection efficiency on the data and MC simulation, evaluated with the  $J/\psi\to\Xi^{-}\bar{\Xi}^{+}$ control sample.

\item {\bf Mass window of \boldmath $\Xi(1530)^0$}: The uncertainty associated with the mass-window requirement on the $\Xi(1530)^0$ candidates is estimated by varying the requirement by $\pm$1 MeV/$c^2$. The largest difference in efficiency between data and MC simulation is taken as the systematic uncertainty.

\item {\bf Fit range}: Any potential bias associated with the choice of fit range is estimated by varying this range by $\pm$10 MeV/$c^2$. The maximum observed difference on the branching fraction is taken as the systematic uncertainty.

\item {\bf Signal shape}:
The uncertainty due to the choice of signal shape is estimated
by changing the MC-simulated shape with a kernel density estimation~\cite{rookeyspdf} of the unbinned signal MC derived shape.
The difference in the measured branching fraction is taken as the systematic uncertainty.

\item {\bf Background shape}:
The uncertainty associated with the use of a Chebychev polynomial function to describe the remaining background events from miscellaneous sources  is estimated by performing alternative fits with a second or fourth-order Chebychev polynomial function.
The difference in the measured branching fraction is taken as the systematic uncertainty.

\item {\boldmath ${\pi^+\pi^- J/\psi}$ \bf background}: The uncertainty associated with the ${\pi^+\pi^- J/\psi}$ background has two components: the impact of the mass-window requirement on $M^{\rm recoil}_{\pi^+\pi^-}$ and the knowledged of the size of background from  $\psi(3686)\to\pi^+\pi^- J/\psi$ with $J/\psi\to\Xi^-\bar\Xi^+$. The uncertainty related to the $M^{\rm recoil}_{\pi^+\pi^-}$ mass window is estimated by changing the requirement by $\pm1$ MeV/$c^2$.
The largest difference on branching fraction is taken as the systematic uncertainty. The uncertainty related to the amount of $\psi(3686)\to\pi^+\pi^- J/\psi$ with $J/\psi\to\Xi^-\bar\Xi^+$ background is estimated by changing the branching fraction of $J/\psi\to\Xi^-\bar\Xi^+$ within one standard deviation. The resulting difference on branching fraction yield is taken as the uncertainty.
The total uncertainty associated with the ${\pi^+\pi^- J/\psi}$ background is calculated by taking the quadratic sum of the two individual contributions.

\item {\bf WCB}: Any possible bias related to WCB is assessed by comparing the signal yields with and without the corresponding component included in the fit. The difference on the branching fraction is found to be negligible.

\item {\bf Angular distribution}: The MC simulation for the signal channels requires as input the angular-distribution parameters of the the baryon pairs. For the $\psi(3686)\to\Xi(1530)^0\bar\Xi(1530)^0$ mode, this uncertainty
 is estimated  by varying the measured $\alpha$ values by $\pm1\sigma$ uncertainty in the MC simulation, and the resulting largest difference compared with the nominal efficiency is taken as the uncertainty.
 For the mode $\Xi(1530)^0\bar\Xi^0$, we set $\alpha=0$ and generate a new MC sample to obtain the detection efficiency, whose difference relative to the nominal efficiency is taken as the uncertainty.
The uncertainty of the MC model for simulating the decays of $\Xi(1530)$, $\Xi$ and $\Lambda$ baryon is estimated by comparing the efficiency with and without considering the angular distribution of baryon pair via {\sc evtgen}~\cite{ref:evtgen}.
The total uncertainty related to the angular distribution is assigned to be the quadratic sum of all individual terms.

\item {\bf Intermediate decay}: The uncertainties due to the branching fractions of the intermediate decays, $\Lambda\to p\pi^-$ and $\Xi^-\to\Lambda\pi^-$, are taken to be 0.8\% and 0.1\%, respectively, from the PDG~\cite{pdg}. The uncertainty of the branching fraction of $\Xi(1530)^0\to\Xi^-\pi^+$ is conservatively taken to be 4.0\%, where we assume $\Xi(1530)$ completely decaying to its $\Xi\pi$ mode~\cite{pdg}.

\item {\boldmath $N_{\psi(3686)}$}: The uncertainties due to the total number of $\psi(3686)$ events ($N_{\psi(3686)}$) are
determined with inclusive hadronic $\psi(3686)$ decays and taken to be 0.7\%~\cite{Npsip}.
\end{itemize}

The total systematic uncertainty in the branching-fraction measurement is obtained by summing the individual contributions in quadrature, which is summarized in Table~\ref{error1}.

\begin{table}[htbp]
\caption{Relative systematic uncertainties on the branching fraction measurements (in \%). }
\begin{center}
\begin{tabular*}{1.02\columnwidth}{@{\extracolsep{\fill}}lcccc}
\hline
\hline
\multirow{2}{2cm}{Source}& \multicolumn{2}{c}{$\Xi(1530)^{0}\bar{\Xi}(1530)^{0}$} & \multicolumn{2}{c}{$\Xi(1530)^{0}\bar{\Xi}^{0}$} \\
        & $\Xi(1530)^0$ & $\bar\Xi(1530)^0$ & $\Xi(1530)^0$ & $\bar\Xi(1530)^0$\\\hline
\hline
Tracking for pion & 1.0 & 1.0 & 1.0 &1.0\\
PID for pion  &  1.0 & 1.0 & 1.0 & 1.0\\
$\Xi^-$ reconstruction & 3.0 & 6.0 & 3.0 & 6.0\\
Decay length $\Lambda$ &0.1 & 0.1& 0.1 & 0.1\\
Decay length $\Xi^-$ & 0.5 &0.4&0.5&0.4\\
Mass window $\Lambda$ & 0.2& 0.4& 0.2 &0.4\\
Mass window $\Xi^-$ & 1.3 &1.3& 1.3& 1.3\\
Mass window $\Xi(1530)^0$ & 0.7 &0.8&1.3&1.5\\
Fit range & 1.1 &0.9&0.6&0.2\\
Signal shape & 1.0 &0.6&2.5&3.4\\
Background shape & 1.5 &1.5&1.2&1.3\\
$\pi^+\pi^- J/\psi$ background & 1.3 &0.8&0.7&0.8\\
WCB & $<0.1$ &$<0.1$&$<0.1$&$<0.1$\\
Angular distribution & 1.7 &2.2&2.2&2.6\\
${\cal{B}}(\Lambda\to p\pi^-)$ & 0.8 &0.8& 0.8&0.8\\
${\cal{B}}(\Xi^-\to \Lambda\pi^-)$ & 0.1 &0.1& 0.1&0.1\\
${\cal{B}}(\Xi(1530)\to\Xi^-\pi^+)$ & 4.0 &4.0&4.0&4.0\\
$N_{\psi(3686)}$ & 0.7 &0.7&0.7&0.7\\
\hline
{\bf Total}      & 6.3 &8.2&6.7&8.9\\
\hline
\hline
\end{tabular*}
\end{center}
\label{error1}
\end{table}

\subsection{Angular distribution}\label{sec:errang}
\vspace{-0.4cm}
The sources of systematic uncertainty for the measurement of the $\alpha$ value include those from the signal yields in different $\cos\theta_{B}$ intervals and those associated with the $\alpha$-value fit. The uncertainties for the signal yield arise from the choice of fit range, the signal and background shapes, the MC model, the level of the WCB and that of the  background size from $\psi(3686)\to\pi^+\pi^- J/\psi$ with $J/\psi\to\Xi^-\bar\Xi^+$.   These uncertainties are estimated with the same method as described in Sec.~\ref{sec:errBR}, where the shifts in the value of $\alpha$ with respect to that from the default fit  are taken as the systematic uncertainties.
The uncertainties associated with the $\alpha$-value fit are assigned as follows.
\begin{itemize}
\item {\bf The \boldmath $\cos\theta_{\rm B}$ \bf binning}: This uncertainty is estimated by comparing the different binning schemes. In the nominal binning scheme, the $\cos\theta_{\rm B}\in$ $[-0.8, 0.8]$ is divided into 8 bins, while an alternative binning scheme divides the $|\rm cos\theta_{\rm B}|\in$ [0, 0.8] into 8 bins. The resulting differences on the $\alpha$ value are
taken as the systematic uncertainties.

\item {\bf The \boldmath $\cos\theta_{\rm B}$ \bf range}: This uncertainty is estimated by repeating the fit after changing the $\cos\theta_{B}$ range to $[-0.7, 0.7]$, with the same bin size as the nominal fit. The differences on the $\alpha$ value are found to be negligible for both tagging modes.
\end{itemize}
All the systematic uncertainties for the $\alpha$ measurement are summarized in Table~\ref{erroralpha}, where the total systematic uncertainty is the quadratic sum of the contributions.

\begin{table}[htbp]
\caption{Systematic uncertainties (absolute) on the measurement of the $\alpha$ value for $\psi(3686)\to\Xi(1530)^{0}\bar{\Xi}(1530)^{0}$.
}
\begin{center}
\renewcommand\arraystretch{1.2} 
\begin{tabular*}{0.9\columnwidth}{@{\extracolsep{\fill}}lcc}
\hline
\hline
Source & $\Xi(1530)^0$ & $\bar\Xi(1530)^0$\\
\hline
Fit range & 0.03 & 0.03\\
Signal shape & 0.05& 0.07\\
Background shape  &  0.03 & 0.04\\
MC model & 0.03& 0.04\\
WCB &0.02 & 0.01\\
$\pi^+\pi^- J/\psi$ background & $\rm Negl.$ & $\rm Negl.$\\

$\cos\theta$ binning & 0.08 &0.03\\
$\cos\theta$ range & $\rm Negl.$ & $\rm Negl.$\\\hline
{\bf Total}      & 0.11 &0.10\\
\hline
\hline
\end{tabular*}
\end{center}
\label{erroralpha}
\end{table}

\section{Results}
\label{sec:result}
Combined branching fractions and $\alpha$ values are calculated with a weighted least-squares method~\cite{leastsq}.
The single-baryon recoil mass method leads to some double counting of the $\Xi(1530)^0\bar\Xi(1530)^0$ final state; MC studies indicate this occurs at a rate of about 10\%, which is taken as a common statistical uncertainty when calculating the combined branching fractions and $\alpha$ value. The systematic uncertainties are weighted to properly account for common and uncommon systematic uncertainties using the method in the Ref.~\cite{memowxf}. The results are summarized in Table~\ref{tab:st0}.  The measured value of $\alpha$ is in agreement with the predictions of Refs.~\cite{ref4}.

To test isospin symmetry, the ratios of the combined branching fractions for the decay $\boldmath{\psi(3686)\to X}$ between neutral and charged-modes, ${\cal{R}}_{\psi(3686)\to X}=\frac{{\cal{B}}^{\rm com}_{\rm neutral}}{{\cal{B}}^{\rm com}_{\rm charged}}$, are calculated by assuming that the common systematic uncertainties between the neutral and charged decay modes
cancel in the measurement.
 The ratio is determined to be $0.59\pm0.03\pm0.06$ for the decay $\boldmath{\psi(3686)\to\Xi(1530)\bar{\Xi}(1530)}$, in significant disagreement with the expectation from isospin symmetry.  The corresponding result for $\boldmath{\psi(3686)\to\Xi(1530)\bar{\Xi}}$ is found to be $0.76\pm0.13\pm0.06$, which is compatible with that expected from isospin symmetry.
 In both cases, the first uncertainty is statistical and the second is systematic.

To test the `12\% rule' in the neutral decay of $\psi\to\Xi(1530)^0\bar{\Xi}^0$, the ratio between ${\cal B}(\psi(3686)\to\Xi(1530)^0\bar{\Xi}^0)$ and ${\cal B}(J/\psi\to\Xi(1530)^0\bar{\Xi}^0)$~\cite{ref200}, ${\cal{Q}}_{\rm neutral}=\frac{{\cal{B}}(\psi(3686)\to \Xi(1530)^0\bar{\Xi}^0)}{{\cal{B}}(J/\psi\to \Xi(1530)^0\bar{\Xi}^0)}$ is calculated by assuming that the correlation between the two measurements is negligible. It is determined to be $(3.31 \pm 1.25\pm0.73)\%$, where the large uncertainty arises from the limited knowledge of  ${\cal{B}}(J/\psi\to \Xi(1530)^0\bar{\Xi}^0)$, where the first and second uncertainties are statistical and systematic, respectively. One observes that
the ratio is suppressed relative to the `12\% rule'. As a comparison, the ratio for its isospin charged mode, ${\cal{Q}}_{\rm charged}=\frac{{\cal{B}}(\psi(3686)\to \Xi(1530)^-\bar{\Xi}^+)}{{\cal{B}}(J/\psi\to \Xi(1530)^-\bar{\Xi}^+)}$, is also calculated with the measured branching fractions~\cite{refjiqp,memowxf} and determined to be $(3. 12\pm0.49\pm0.20)\%$.
Hence the ratios $\cal{Q}_{\rm neutral}$ and $\cal{Q}_{\rm charged}$ are compatible in value, and both of them disfavor the `12\% rule'.

\section{Summary}
\label{sec:summary}
\vspace{-0.4cm}
In summary, using $(448.1\pm2.9)\times 10^{6}$ $\psi(3686)$ events collected with the BESIII detector at the BEPCII,
we have made the first observation of $\psi(3686)\to\Xi(1530)^{0}\bar{\Xi}(1530)^{0}$ and $\Xi(1530)^{0}\bar{\Xi}^0$ decays by a single-baryon tagging technique.
 The corresponding branching fractions are measured to be $(6.77\pm0.14\pm0.39)\times10^{-5}$ and $(0.53\pm0.04\pm0.03)\times10^{-5}$, respectively.
The branching fraction for $\psi(3686)$$\to$$\Xi(1530)^{0}\bar{\Xi}(1530)^{0}$ measured in this work is consistent with the result reported by the CLEO collaboration~\cite{cleo1530}, and the precision is significantly improved by about 40 times. The observation of the  $\psi(3686)\to\Xi(1530)^{0}\bar{\Xi}^0$ decay mode indicates that SU(3) flavor
symmetry is broken in $\psi(3686)$ decays, which further validates the generality of SU(3) flavor symmetry breaking.
It is also found that isospin symmetry is violated in the decay $\psi(3686)\to\Xi(1530)\bar{\Xi}(1530)$, but conserved within $1.5\sigma$ of the expectation for the decay $\psi(3686)\to\Xi(1530)\bar{\Xi}$. The `12\% rule' is tested in the decay $\psi\to \Xi(1530)\bar{\Xi}$ and found to be highly violated.
The measured angular distribution parameter $\alpha$ for $\psi(3686)\to\Xi(1530)^{0}\bar{\Xi}(1530)^{0}$ decay, $0.32\pm0.19\pm0.07$, agrees with the charged mode~\cite{memowxf}, as well as the
theoretical predictions~\cite{ref4} within one standard deviation.

~~~~~~~~~~~~~
\begin{acknowledgements}
\label{sec:acknowledgement}
\vspace{-0.4cm}
The BESIII collaboration thanks the staff of BEPCII and the IHEP computing center for their strong support. This work is supported in part by National Key Basic Research Program of China under Contracts Nos. 2020YFA0406300, 2020YFA0406400; National Natural Science Foundation of China (NSFC) under Contracts Nos. 11605042, 11625523, 11635010, 11735014, 11822506, 11835012, 11875122, 11935015, 11935016, 11935018, 11905236, 11961141012, 12022510, 12025502, 12035009, 12035013, 1204750, 12075107, 12061131003; the Chinese Academy of Sciences (CAS) Large-Scale Scientific Facility Program; Joint Large-Scale Scientific Facility Funds of the NSFC and CAS under Contracts Nos. U1732263, U1832207; CAS Key Research Program of Frontier Sciences under Contract No. QYZDJ-SSW-SLH040; 100 Talents Program of CAS; INPAC and Shanghai Key Laboratory for Particle Physics and Cosmology; ERC under Contract No. 758462; European Union Horizon 2020 research and innovation programme under Contract No. Marie Sklodowska-Curie grant agreement No 894790; German Research Foundation DFG under Contracts Nos.~443159800, Collaborative Research Center CRC 1044, FOR 2359, GRK 214; Istituto Nazionale di Fisica Nucleare, Italy; Ministry of Development of Turkey under Contract No.~DPT2006K-120470; National Science and Technology fund; Olle Engkvist Foundation under Contract No. 200-0605; STFC (United Kingdom); The Knut and Alice Wallenberg Foundation (Sweden) under Contract No. 2016.0157; The Royal Society, UK under Contracts Nos.~DH140054, DH160214; The Swedish Research Council; U. S. Department of Energy under Contracts Nos. DE-FG02-05ER41374, DE-SC-0012069; Excellent Youth Foundation of Henan Province under Contracts No. 212300410010;
The youth talent support program of Henan Province under Contracts No. ZYQR201912178; Program for Innovative Research Team in University of Henan Province under Contracts No. 19IRTSTHN018; The Fundamental Research Funds for the Central Universities under Grants No. lzujbky-2021-sp24.

\end{acknowledgements}

\end{document}